\documentclass[11pt]{article}
\usepackage{mathrsfs}
\usepackage{amsmath}
\usepackage{amsfonts}
\usepackage{amssymb, amsmath, cite}
\usepackage{color}

\newtheorem{theorem}{Theorem}
\newtheorem{lemma}{Lemma}

\newcommand\be{\begin{equation}}
\newcommand\ee{\end{equation}}
\newcommand\ber{\begin{eqnarray}}
\newcommand\eer{\end{eqnarray}}
\newcommand\berr{\begin{eqnarray*}}
\newcommand\eerr{\end{eqnarray*}}
\newcommand\bea{\begin{eqnarray}}
\newcommand\eea{\end{eqnarray}}

\newcommand\bfR{\mathbb{R}}\newcommand\dd{\mbox{d}}\newcommand\ii{\mbox{i}}
\newcommand\re{\mathrm{e}}
\newcommand\ri{\mathrm{i}}
\newcommand{\ud}{\mathrm{d}}
\newcommand{\nn}{\nonumber}

\newcommand{\vep}{\varepsilon}

\title{Topologically Stratified Energy Minimizers\\in a Product Abelian Field Theory}

\author{Xiaosen Han\\Institute of Contemporary Mathematics\\School of Mathematics\\Henan University\\
Kaifeng, Henan 475004, PR China\\ \\
Yisong Yang\\Department of Mathematics\\Polytechnic School of Engineering\\ New York University\\Brooklyn, New York 11201, USA
\\  \&\\NYU-ECNU
Institute of Mathematical Sciences\\New York University - Shanghai\\3663 North Zhongshan Road, Shanghai 200062, PR China}

\date{}

\begin{document}
\maketitle
\begin{abstract}
We study a recently developed product Abelian gauge field theory by Tong and Wong hosting magnetic impurities.
We first obtain a necessary and sufficient condition for the existence of a unique solution
realizing such impurities in the form of multiple vortices. We next reformulate the theory into an extended model
that allows the coexistence of vortices and anti-vortices. The two Abelian gauge fields in the model induce two species of magnetic vortex-lines resulting from
$N_s$ vortices and $P_s$ anti-vortices ($s=1,2$) realized as the zeros and poles of two complex-valued Higgs fields, respectively. An
existence theorem is established for the governing equations over a compact Riemann surface $S$ which states that a solution with prescribed $N_1, N_2$ vortices
and $P_1,P_2$ anti-vortices of two designated species exists if and only if the inequalities
\[
\left|N_1+N_2-(P_1+P_2)\right|<\frac{|S|}{\pi},\quad \left|N_1+2N_2-(P_1+2P_2)\right|<\frac{|S|}{\pi},
\]
hold simultaneously, which give bounds for the `differences' of the vortex and anti-vortex numbers in terms of the total surface area of $S$. The minimum energy of these solutions
is shown to assume the explicit value
\[
E= 4\pi (N_1+N_2+P_1+P_2),
\]
given in terms of several topological invariants, measuring the total tension of the vortex-lines.

\medskip

\end{abstract}

\section{Introduction}
\setcounter{equation}{0}

It is well known that the simplest and most important quantum field theory model is the Abelian Higgs model \cite{JT,Ryder} which embodies an electromagnetic gauge field and
spontaneously broken symmetry and allows mass generation through the Higgs mechanism. In the temporal gauge, its static limit gives rise to the classical Ginzburg--Landau
theory for superconductivity \cite{GL} so that in its two dimensional setting mixed-state configurations known as the Abrikosov vortices \cite{Ab} can be rigorously constructed
\cite{Br,JT,T1,T2,WY}. Inspired by the gauged sigma model of Schroers \cite{Sch1,Sch2}, the classical Abelian Higgs model is extended in \cite{Y1,Y2} to allow the coexistence
of vortices and anti-vortices. This extended model is also shown to generate cosmic strings and anti-strings when gravitation is switched on by the Einstein equations which give
rise to curvature and mass concentrations essential for matter accretion in the early universe
\cite{Kibble1,Kibble2,Vi,VS,Yob1,Yob2}. In order to understand the topological contents of such an extended Abelian Higgs model, a reformulation of it is carried in
\cite{SSY} in the context of
a complex line bundle over a compact Riemann surface $S$ as in \cite{Br,Ga,N1,N2}. In a sharp and interesting contrast with the Abelian Higgs model where vortices are topologically
characterized by the first Chern class, the vortices and anti-vortices in the extended Abelian Higgs model \cite{Y1,Y2} are characterized jointly and elegantly \cite{SSY} by the first Chern class of the
line bundle and the Thom class \cite{DS} of the associated dual bundle. In the former case, there are only finitely many minimum energy values which can be attained due to the fact that
the total number of vortices is confined by the total  area $|S|$ of the two-surface $S$ where vortices reside. In the latter case, however, the confinement is made instead to the difference of
the numbers of vortices and anti-vortices, but the minimum energy is proportional to the sum of these numbers. Hence the possible minimum energy values becomes an explicitly determined
infinite sequence as in the situation of vortices over a non-compact surface in the classical Abelian Higgs theory \cite{JT,T1,T2}.

In a recent interesting work of Tong and Wong \cite{TW}, a product Abelian gauge field theory is formulated to include magnetic impurities in the form of an extra gauge-matter sector. This gauge-matter
sector is not treated as a background source but as a fully coupled sector. In other words, this is a product Abelian gauge field theory with two complex Higgs fields. It is shown in \cite{TW}
that, like in the classical Abelian Higgs model, the new product model allows a BPS (after Bogomol'nyi \cite{Bo} and Prasad--Sommerfield \cite{PS}) reduction, hence a construction of magnetic
vortices as in \cite{JT}. The present paper aims to enrich our understanding of Abelian (magnetic) vortices by achieving two goals. The first is to extend the product Abelian gauge field theory of Tong and Wong \cite{TW} using the ideas in
\cite{Sch1,Sch2,Y1,Y2} into a new product field theory that allows the coexistence of two species of vortices and anti-vortices. The second is to establish an existence theorem for such
vortices of beautiful topological characteristics. For clarity and simplicity, the underlying domain for the vortices to live is assumed to be a compact Riemann surface, as in \cite{SSY}.

In order to put our study in an appropriate perspective, we shall first present a reformulation of the Tong--Wong model \cite{TW} in terms of
a complex line bundle over a compact Riemann surface. In such a context, we show that a Bradlow type bound or limit appears as in the
Abelian Higgs theory for the existence of multiple vortices \cite{Br,Ga,N1,N2,WY}, which is a preparation for our work regarding the extended model.

An outline of the rest of the paper is as follows.
In Section 2, we present the Tong--Wong theory and our extended product Abelian gauge field theory, in their static limits.
 We describe in detail the field-theoretical properties of the extended theory and derive its BPS equations. We then state our main existence theorems for the existence of multiple vortices in the Tong--Wong theory and
 for the coexistence of multiple vortices and anti-vortices, of two species. In Section 3,
 we convert the BPS equations into systems of nonlinear elliptic equations, state the main existence theorems
 in terms of these equations, and carry out some preliminary discussion. In Section 4, we establish the existence and uniqueness
 theorem for the Tong--Wong multiple vortex solutions by calculus of variations.
In Section 5, we prove the existence theorem for the vortex and anti-vortex solutions
of our extended model by using a Leray--Schauder fixed-point theorem argument \cite{GT}
under a necessary and sufficient condition. In Section 6, we explicitly compute the (minimum) energy of a vortex and anti-vortex solution and show that such energy arises topologically and is proportional to the sum of vortex and anti-vortex numbers of two species.
In Section 7, we make some concluding remarks regarding coexisting vortices and anti-vortices in the extended model.

\section{Energy functionals, BPS reductions, and existence theorems}
\setcounter{equation}{0}

Let $L$ be complex Hermitian line bundle over a Riemann surface $S$. Use $q,p$ to denote two sections $L\to S$ and $Dq,Dp$ the connections induced from
the real-valued connection 1-forms $\hat{A},\tilde{A}$, respectively, so that
\be
Dq=\dd q-\ii (\hat{A}-\tilde{A})q,\quad Dp=\dd p-\ii \tilde{A} p.
\ee
 Using $*$ to denote the usual Hodge dual operating on differential forms, the energy density of the
Tong--Wong model \cite{TW} for a product
Abelian Higgs theory implementing magnetic impurities may be rewritten as
\bea\label{x1.2}
{\cal E}&=&\frac12 *(\hat{F}\wedge *\hat{F})+\frac12*(\tilde{F}\wedge *\tilde{F})+*(Dq\wedge * \overline{Dq})+*(Dp\wedge * \overline{Dp})\nn\\
&&+\frac12(1-|q|^2)^2+\frac12([1-|q|^2]+[|p|^2-1])^2,
\eea
where $\hat{F}=\dd\hat{A},\tilde{F}=\dd\tilde{A}$ are curvature 2-forms, which recovers the classical Ginzburg--Landau model
\cite{GL} when impurities are switched off by setting
\be
\tilde{A}=0,\quad p=1.
\ee

Note also that there holds the identity
\be
Dq\wedge *\overline{Dq}+(*Dq)\wedge \overline{Dq}=(Dq\pm\ii *Dq)\wedge *\overline{(Dq\pm\ii * Dq)}
\pm\ii (Dq\wedge \overline{Dq}-(*Dq)\wedge(*\overline{Dq})).
\ee
Thus we get
\bea\label{x2.5}
{\cal E}&=&\frac12\left|\hat{F}\mp *(1-|q|^2)\right|^2+\frac12\left|\tilde{F}\pm(*[1-|q|^2]+*[|p|^2-1])\right|^2\nn\\
&&\pm *\hat{F}(1-|q|^2)\mp *\tilde{F}([1-|q|^2]+[|p|^2-1])\nn\\
&&+|Dq\pm\ii *Dq|^2+|Dp\pm\ii * Dp|^2\nn\\
&&\pm\frac{\ii}2 (Dq\wedge \overline{Dq}-(*Dq)\wedge(*\overline{Dq}))\pm\frac{\ii}2 (Dp\wedge \overline{Dp}-(*Dp)\wedge(*\overline{Dp})).
\eea

On the other hand, with the current densities
\be
J(q)=\frac{\ii}2(q\overline{Dq}-\overline{q} Dq),\quad J(p)=\frac{\ii}2(p\overline{Dp}-\overline{p} Dp),
\ee
we have
\bea
\dd J(q)&=&-(\hat{F}-\tilde{F})|q|^2+\frac{\ii}2(Dq\wedge\overline{Dq}-*Dq\wedge *\overline{Dq}),\label{x2.7}\\
\dd J(p)&=&-\tilde{F}|p|^2+\frac{\ii}2(Dp\wedge\overline{Dp}-*Dp\wedge *\overline{Dp}).\label{x2.8}
\eea

Inserting (\ref{x2.7}) and (\ref{x2.8}) into (\ref{x2.5}), we have
\bea\label{x2.9}
{\cal E}&=&\frac12\left|\hat{F}\mp *(1-|q|^2)\right|^2+\frac12\left|\tilde{F}\pm(*[1-|q|^2]+*[|p|^2-1])\right|^2\nn\\
&&+|Dq\pm\ii *Dq|^2+|Dp\pm\ii * Dp|^2\nn\\
&&\pm *(\hat{F}-\tilde{F})\pm *\tilde{F}\pm*\dd J(q)\pm*\dd J(p),
\eea
which leads to the topological energy lower bound
\be\label{x2.10}
E=\int_S {\cal E}*1\geq\left|\int_S \hat{F}\right|.
\ee
From the form of (\ref{x2.9}) it is clear that the lower bound in (\ref{x2.10}) is attained by the solutions of the BPS equations
\bea
\hat{F}&=&\pm *(1-|q|^2),\label{x2.11}\\
\tilde{F}&=&\mp *([1-|q|^2]+[|p|^2-1]),\label{x2.12}\\
Dq\pm \ii *Dq&=&0,\label{x2.13}\\
Dp\pm \ii *Dp&=&0,\label{x2.14}
\eea
as derived by Tong and Wong in \cite{TW}.

The equations of motion of (\ref{x1.2}) are
\bea
D*Dq&=&-(1-|q|^2)q-([1-|q|^2]+[|p|^2-1])q,\label{x2.15}\\
\dd*\hat{F}&=&\ii(\overline{q}Dq-q\overline{Dq}),\label{x2.16}\\
D*Dp&=&([1-|q|^2]+[|p|^2-1])p,\label{x2.17}\\
\dd *\tilde{F}&=&-\ii(\overline{q}Dq-q\overline{Dq})+\ii (\overline{p}Dp-p\overline{Dp}),\label{x2.18}
\eea
which contain (\ref{x2.11})--(\ref{x2.14}) as its first integral and may be viewed as
a reduced form of (\ref{x2.15})--(\ref{x2.18}). These reduced first-order equations are often referred to as the BPS equations after Bogomol'nyi \cite{Bo} and Prasad--Sommerfield \cite{PS} who pioneered the
idea of such reduction for the classical Yang--Mills--Higgs equations. When the upper sign is taken, the system is said to be self-dual; the lower, anti-self-dual.
It may also be checked that the self-dual and anti-self-dual cases are related to each other through the transformation
\be
\hat {A} \rightleftharpoons -\hat{A},\quad\tilde{A}\rightleftharpoons- \tilde{A},\quad q \rightleftharpoons\overline{q},\quad p\rightleftharpoons\overline{p}.
\ee
Hence, in the sequel, we will only consider the self-dual situation.

The structure of (\ref{x2.13}) and (\ref{x2.14}) indicates that the zeros of $q,p$ are isolated and of integer multiplicities which may be assumed to be
\be\label{xz}
{\cal Z}(q)=\{z_{1,1},\dots,z_{1,N_1}\},\quad {\cal Z}(p)=\{z_{2,1},\dots,z_{2,N_2}\},
\ee
where for convenience a zero of multiplicity $m$ is counted as $m$ zeros in the zero set. The quantities $\frac1{2\pi}\int (\hat{F}-\tilde{F})$ and $\frac1{2\pi}\int \tilde{F}$ are the first Chern numbers induced from the connections $\hat{A}-\tilde{A}$ and $\tilde{A}$ over
 $L\to S$ which are determined by the numbers of zeros, $N_1$ and $N_2$, by the formulas
 \bea
 \frac1{2\pi}\int_S (\hat{F}-\tilde{F})&=& N_1,\label{221}\\
\frac1{2\pi}\int_S\tilde{F}&=& N_2,\label{222}
\eea
respectively.

Regarding the Tong--Wong BPS equations (\ref{x2.11})--(\ref{x2.14}), here is our existence theorem.

\begin{theorem}\label{thm2.1}
For the BPS system consisting of equations (\ref{x2.11})--(\ref{x2.14}) over a compact Riemann surface $S$ with canonical total area
$|S|$ governing two connection 1-forms $\hat{A},\tilde{A}$ and two cross sections $q,p$ with the prescribed sets of zeros
given  in (\ref{xz}), there exists a solution to realize
these sets of zeros if and only if $N_1$ and $N_2$ satisfy the bound
\be\label{Bradlow}
N_1+2N_2<\frac{|S|}{2\pi}.
\ee
Such a solution carries a minimum
energy of the form
\be\label{xE}
E= 2\pi (N_1+N_2),
\ee
and is unique up to gauge transformations.
\end{theorem}

The condition stated in (\ref{Bradlow}) is analogous to the so-called Bradlow bound \cite{BM,Dem,Nasir} in
the classical Abelian Higgs model \cite{Br,Ga,WY} which was actually deduced earlier by Noguchi \cite{N1,N2}.

Next, following \cite{Y1,Y2} based on the idea of gauged sigma model, we show that we may extend the Tong--Wong model \cite{TW} to accommodate vortices and anti-vortices by considering the modified energy density
\bea
{\cal E}&=&\frac12 *(\hat{F}\wedge *\hat{F})+\frac12*(\tilde{F}\wedge *\tilde{F})+\frac4{(1+|q|^2)^2}*(Dq\wedge * \overline{Dq})+\frac4{(1+|p|^2)^2}*(Dp\wedge * \overline{Dp})\nn\\
&&+2\left(\frac{1-|q|^2}{1+|q|^2}\right)^2
+2\left(\frac{1-|q|^2}{1+|q|^2}+\frac{|p|^2-1}{1+|p|^2}\right)^2. \label{x1.4}
\eea

In fact, let $\phi$ and $\psi$ be two $S^2$-valued scalar fields. Fix ${\bf n}=(0,0,1)\in S^2$ and define the vacuum manifold of the model to be
\be
{\bf n}\cdot\phi=0,\quad {\bf n}\cdot\psi=0.
\ee
Then we modify (\ref{x1.2}) into the form
\be\label{xx1.2}
{\cal E}=\frac12 *(\hat{F}\wedge *\hat{F})+\frac12*(\tilde{F}\wedge *\tilde{F})+|D\phi|^2+|D\psi|^2+
2({\bf n}\cdot\phi)^2+2({\bf n}\cdot[\phi-\psi])^2,
\ee
where
\be
D\phi=\dd \phi-({\bf n}\times\phi)(\hat{A}-\tilde{A}),\quad D\psi=\dd \psi-({\bf n}\times\psi)\tilde{A}.
\ee
Use the stereographic projection from the south pole $-{\bf n}$ of $S^2$ to represent $\phi$ and $\psi$ by complex-valued functions $q$ and $p$,
respectively, so that
\be\label{2.29}
\phi=\left(\frac{2\Re\{q\}}{1+|q|^2},\frac{2\Im\{q\}}{1+|q|^2},\frac{1-|q|^2}{1+|q|^2}\right),\quad \psi=\left(\frac{2\Re\{p\}}{1+|p|^2},\frac{2\Im\{p\}}{1+|p|^2},\frac{1-|p|^2}{1+|p|^2}\right),
\ee
where $\Re\{c\}$ and $\Im\{c\}$ denote the real and imaginary parts of a complex number $c$. Inserting (\ref{2.29}) into (\ref{xx1.2}), we arrive at (\ref{x1.4}).

It is interesting to observe that (\ref{x1.2}) is recovered from (\ref{x1.4}) when taking the limit $|q|\to 1,|p|\to 1$ in the
denominators $1+|q|^2$ and $1+|p|^2$ of (\ref{x1.4}).
The Euler--Lagrange equations of the energy density are found to be
\bea
D*\left(\frac{Dq}{(1+|q|^2)^2}\right)&=&\frac1{(1+|q|^2)^3}(Dq\wedge *\overline{Dq})+2*\left(\frac{1-|q|^2}{(1+|q|^2)^3}\right)q\nn\\
&& +2*\left(\frac{1-|q|^2}{1+|q|^2}+\frac{|p|^2-1}{1+|p|^2}\right)\left(\frac{1-|q|^2}{(1+|q|^2)^2}\right)q,\label{x1.5}\\
\dd*\hat{F}&=&4\ii \frac{(\overline{q}Dq-q\overline{Dq})}{(1+|q|^2)^2},\label{x1.6}\\
D*\left(\frac{Dp}{(1+|p|^2)^2}\right)&=&\frac1{(1+|p|^2)^3}(Dp\wedge *\overline{Dp})\nn\\
&&+2*\left(\frac{1-|q|^2}{1+|q|^2}+\frac{|p|^2-1}{1+|p|^2}\right)\left(\frac{|p|^2-1}{(1+|p|^2)^2}\right)p,\label{x1.7}\\
\dd*\tilde{F}&=&-4\ii \frac{(\overline{q}Dq-q\overline{Dq})}{(1+|q|^2)^2}+4\ii \frac{(\overline{p}Dp-p\overline{Dp})}{(1+|p|^2)^2},\label{x1.8}
\eea
which appear rather complicated and intractable. In order to obtain interesting solutions of these equations, we follow \cite{TW,Y1} to pursue a BPS reduction.

Introduce the current densities
\be
J(q)=\frac{\ii}{1+|q|^2}(q\overline{Dq}-\overline{q} Dq),\quad J(p)=\frac{\ii}{1+|p|^2}(p\overline{Dp}-\overline{p} Dp).
\ee
Then we have
\bea
K(q)&=&\dd J(q)\nn\\
&=&-\frac{2 |q|^2}{1+|q|^2}(\hat{F}-\tilde{F})+\ii\left(\frac{Dq\wedge \overline{Dq}-*Dq\wedge *\overline{Dq}}{(1+|q|^2)^2}\right),\\
K(p)&=&\dd J(p)\nn\\
&=&-\frac{2|q|^2}{1+|q|^2}\tilde{F}+\ii\left(\frac{Dp\wedge \overline{Dp}-*Dp\wedge *\overline{Dp}}{(1+|p|^2)^2}\right).
\eea

So, with $|Dq|^2=*(Dq\wedge *\overline{Dq})$, etc, we arrive at the decomposition
\bea
{\cal E}&=&\frac12\left|\hat{F}\mp 2*\left(\frac{1-|q|^2}{1+|q|^2}\right)\right|^2+\frac12\left|\tilde{F}\pm \left(2*\left(\frac{1-|q|^2}{1+|q|^2}\right)
+2*\left(\frac{|p|^2-1}{1+|p|^2}\right)\right)\right|^2\nn\\
&&+\frac2{(1+|q|^2)^2}|Dq\pm\ii *Dq|^2+\frac2{(1+|p|^2)^2}|Dp\pm\ii *Dp|^2\nn\\
&&\pm 2*(\hat{F}-\tilde{F})\pm 2* K(q)\pm 2 *\tilde{F}\pm 2*K(p). \label{x1.13}
\eea

The quantities
$\frac1{4\pi}\int K(q)$ and $\frac1{4\pi}\int K(p)$ are the Thom classes over $L^*\to S$, respectively \cite{SSY}.
Thus,  the sum
\be
\tau =  2\hat{F}+2K(q)+2K(p) \label{x1.14}
\ee
is a topological density which leads to the topological energy lower bound
\be
E=\int_M {\cal E}\,*1\geq\left|\int_M\tau\right|, \label{x1.15}
\ee
measuring the tension \cite{CY,E1,E2,E3,E4} of the vortex-lines, so that the lower bound is saturated when the quartet $(q,p,\hat{A},\tilde{A})$ satisfies the equations
\bea
Dq\pm\ii *Dq&=&0,\label{x1.17}\\
Dp\pm\ii *Dp&=&0,\label{x1.18}\\
\hat{F}&=&\pm 2*\left(\frac{1-|q|^2}{1+|q|^2}\right),\label{x1.19}\\
\tilde{F}&=&\mp \left(2*\left(\frac{1-|q|^2}{1+|q|^2}\right)+2*\left(\frac{|p|^2-1}{1+|p|^2}\right)\right).\label{x1.20}
\eea

It may directly be checked that (\ref{x1.17})--(\ref{x1.20}) imply (\ref{x1.5})--(\ref{x1.8}). In other words,  (\ref{x1.17})--(\ref{x1.20}) may be regarded as a reduction of the system of equations
 (\ref{x1.5})--(\ref{x1.8}).

From (\ref{x1.17}) and (\ref{x1.18}), we know \cite{JT,Y1,Y2} that the zeros and poles of the sections $q,p$ are
isolated and possess integer multiplicities. For simplicity, we may denote the sets of zeros and poles of $q,p$  by
\bea
&& {\cal Z}(q)=\{z_{1,1}',\dots,z_{1,N_1}'\},\quad {\cal P}(q)=\{z_{1,1}'',\dots,z_{1,P_1}''\},\label{Zq}\\
&& {\cal Z}(p)=\{z_{2,1}',\dots,z_{2,N_2}'\},\quad {\cal P}(p)=\{z_{2,1}'',\dots,z_{2,P_2}''\},\label{Zp}
\eea
respectively, so that the associated multiplicities of the zeros and poles are naturally counted by their repeated appearances
in the above collections of points.

If we interpret $*\hat{F}$ as a magnetic or vorticity field, (\ref{x1.19}) indicates that it attains its maximum $*\hat{F}=2$ at the zeros
and minimum $*\hat{F}=-2$ at the poles of $q$. Thus, the zeros and poles of $q$ may be viewed as centers of vortices and anti-vortices.
In other words, we may identify the zeros and poles of $q$ as the locations of vortices and anti-vortices generated from the connection 1-form
$\hat{A}$. Similarly, the zeros and poles of $p$ may be interpreted as vortices and anti-vortices generated from the connection 1-form
$\hat{A}+\tilde{A}$. Therefore, in what follows, the zeros and poles of $q,p$ are interchangeably
and generically referred to as the vortices and anti-vortices of a
solution configuration $(\hat{A},\tilde{A},q,p)$.

Here is our   existence theorem for the BPS equations (\ref{x1.17})--(\ref{x1.20}).
\begin{theorem}\label{mainthm}
Consider the BPS system consisting of equations (\ref{x1.17})--(\ref{x1.20}) of the energy density (\ref{x1.4})
formulated over a complex Hermitian line bundle $L$ over a compact Riemann surface $S$ with canonical total area
$|S|$ governing two connection 1-forms $\hat{A},\tilde{A}$ and two cross sections $q,p$ and comprising a reduction of
the Euler--Lagrange equations (\ref{x1.5})--(\ref{x1.8}). For the prescribed sets of zeros and poles for the fields $q$ and $p$
given respectively in (\ref{Zq}) and (\ref{Zp}), the coupled equations (\ref{x1.17})--(\ref{x1.20}) have a solution to realize
these sets of zeros and poles, if and only if the inequalities
\bea
\left|N_1+N_2-(P_1+P_2)\right|&<&\frac{|S|}{\pi},\label{C1}\\
 \left|N_1+2N_2-(P_1+2P_2)\right|&<&\frac{|S|}{\pi},\label{C2}
\eea
regarding the total numbers of zeros and poles are fulfilled simultaneously. Moreover, such a solution carries a minimum
energy of the form
\be\label{xE}
E= 4\pi (N_1+N_2+P_1+P_2),
\ee
which is seen to be stratified topologically by the Chern and Thom classes of the line bundle $L$ and its dual respectively.
In particular, in terms of energy, zeros (vortices) and poles (anti-vortices) of $q,p$ contribute equally.
\end{theorem}

It is interesting to note that the inequalities (\ref{C1}) and (\ref{C2}) imply that the differences of
vortices and anti-vortices must stay within suitable ranges to ensure the existence of a solution:
\ber
  \Big|N_1-P_1\Big|&<& \frac{3|S|}{\pi}, \label{b3}\\
  \Big|N_2-P_2\Big|&<& \frac{2|S|}{\pi}.\label{b4}
   \eer
  However, it may be checked that the conditions (\ref{b3}) and (\ref{b4}) do not lead to (\ref{C1}) and (\ref{C2}). The latter may be called
  the difference of total numbers of vortices and anti-vortices and the difference of `weighted total numbers' of vortices and anti-vortices.
We note that (\ref{b3}) and (\ref{b4}) give the upper bounds of the total `magnetic fluxes'
\bea
\int_S (\hat{F}-\tilde{F})&=&2\pi(N_1-P_1),\\
\int_S \tilde{F}&=&2\pi (N_2-P_2),
\eea
generated by the `magnetic fields' $\hat{F}-\tilde{F}$
and $\tilde{F}$, respectively, which may be compared with (\ref{221}) and (\ref{222}) for the fluxes of the Tong--Wong model \cite{TW}.
See \S 6 for details of calculation.

\section{Governing elliptic equations and basic properties}
\setcounter{equation}{0}
\setcounter{theorem}{0}

To proceed, we set
\be\label{x3.1}
u=\ln|q|^2,\quad v=\ln|p|^2,
\ee
in (\ref{x2.11})--(\ref{x2.14}). Thus
by \cite{JT,Y1,Y2} we are led to the following equivalent governing elliptic equations
\bea
\Delta u&=&4(\re^u-1)-2(\re^v-1)+4\pi\sum_{z\in{\cal Z}(q)}\delta_z,\label{x3.2}\\
\Delta v&=&-2(\re^u-1)+2(\re^v-1)+4\pi\sum_{z\in{\cal Z}(p)}\delta_z,\label{x3.3}
\eea
where  $\Delta$  is the Laplace--Beltrami operator on $(S, g)$ defined by
 \be \Delta u=\frac{1}{\sqrt{g}}\partial_j(g^{jk}\sqrt{g}\partial_ku), \ee
and $\delta_{z}$ denotes the Dirac measure concentrated at the point $z\in S$ with respect to the Riemannian metric $g$ over $S$.

In what follows, we use $\dd\Omega_g$ to denote the canonical surface element and $|S|$ the associated total area of the Riemann surface  $(S, g)$.

The results stated in Theorem \ref{thm2.1} are contained in the following theorem concerning
the coupled elliptic equations (\ref{x3.2}) and (\ref{x3.3}).

\begin{theorem}\label{thm3.1}
The system of equations consisting of (\ref{x3.2}) and (\ref{x3.3}) has a solution if only only if

\be N_1+2N_2<\frac{|S|}{2\pi}.\label{z0}\ee
Moreover,
if a solution exists, it must be unique and satisfies the quantization conditions
\bea
\int_S (1-\re^u)\,\ud \Omega_g&=&2\pi (N_1+N_2),\label{z01}\\
\int_S (1-\re^v)\,\ud \Omega_g&=&2\pi(N_1+2N_2).\label{z02}
\eea
\end{theorem}

Similarly, setting (\ref{x3.1}) in  (\ref{x1.17})--(\ref{x1.20}), we obtain
\ber
\Delta u&=&\frac{8 (\re^u-1)}{\re^u+1}-\frac{4(\re^v-1)}{\re^v+1}+4\pi\sum\limits_{z\in{\cal Z}(q)}\delta_{z}-4\pi\sum\limits_{z\in{\cal P}(q)}\delta_{z}, \label{b1}\\
\Delta v&=&-\frac{4(\re^u-1)}{\re^u+1}+\frac{4(\re^v-1)}{\re^v+1}+4\pi\sum\limits_{z\in{\cal Z}(p)}\delta_{z}-4\pi\sum\limits_{z\in{\cal P}(p)}\delta_{z}. \label{b2}
\eer

Regarding the equivalently reduced equations (\ref{b1}) and (\ref{b2}) from (\ref{x1.17})--(\ref{x1.20}), we have

\begin{theorem}\label{thb1}
The  coupled equations  \eqref{b1} and \eqref{b2} admit a solution $(u,v)$ with
 the prescribed   sets $\mathcal{Z}(q), \mathcal{P}(q), \mathcal{Z}(p), \mathcal{P}(p)$ in  $S$ specified in (\ref{Zq}) and (\ref{Zp})
 if and only if the  inequalities
 (\ref{C1}) and (\ref{C2})
  are satisfied simultaneously.
Moreover, for the solution to the equations   \eqref{b1} and \eqref{b2} obtained above, there hold the quantized integrals
  \ber
   \int_S \frac{1-\re^u}{1+\re^u}\,\ud\Omega_g=\pi\left(N_1-P_1+N_2-P_2\right),\label{b4a}\\
      \int_S \frac{1-\re^v}{1+\re^v}\,\ud\Omega_g=\pi\left(N_1-P_1+2(N_2-P_2)\right)\label{b4b}.
  \eer
\end{theorem}

For convenience, we first need to take care of the Dirac distributions by subtracting suitable background functions. To do so,
we let $u_0^1, u_0^2,v_0^1, v_0^2$  be the normalized solutions of the equations that determine the source functions
arising from the sets ${\cal Z}(q),{\cal P}(q), {\cal Z}(p),{\cal P}(p)$, respectively. For instance, $u_0^1$ is the unique solution \cite{Aubin} to
\be
\Delta u_0^1=-\frac{4\pi N_1}{|S|}+4\pi\sum\limits_{z\in{\cal Z}(q)}\delta_z, \quad \int_Su_0^1\,\ud\Omega_g=0.\label{b5}
\ee

Set
$u=u_0^1+U,  v=v_0^1+V.$
Then we can rewrite   \eqref{x3.2} and \eqref{x3.3} as
\ber
 \Delta U&=&4(\re^{u_0^1+U}-1)- 2(\re^{v_0^1+V}-1)+\frac{4\pi N_1}{|S|}, \label{z3}\\
 \Delta V&=&-2(\re^{u_0^1+U}-1)+2(\re^{v_0^1+V}-1)+\frac{4\pi N_2}{|S|}.\label{z4}
\eer

We first show the necessity of the condition \eqref{z0}.
If there is a solution of \eqref{z3}--\eqref{z4}, integration of which over $S$ gives
\ber
 \int_S\re^{u_0^1+U}\ud \Omega_g=|S|-2\pi(N_1+N_2)\equiv a_1>0,\label{zz1}\\
 \int_S\re^{v_0^1+V}\ud \Omega_g=|S|-2\pi(N_1+2N_2)\equiv a_2>0.\label{zz2}
\eer
Then we see that the condition  \eqref{z0} is necessary to ensure the existence of a solution to
the system \eqref{z3}--\eqref{z4}.  The quantized integrals \eqref{z01}--\eqref{z02} follow from
\eqref{zz1}--\eqref{zz2}.

Let
$u=u_0^1-u_0^2+U,  v=v_0^1-v_0^2+V.$
 Then the  equations   \eqref{b1} and \eqref{b2} can be rewritten as
\ber
 \Delta U&=&8  f(u_0^1, u_0^2, U)- 4f(v_0^1, v_0^2, V)+\frac{4\pi(N_1-P_1)}{|S|}, \label{b7}\\
 \Delta V&=&-4 f(u_0^1, u_0^2, U)+4 f(v_0^1, v_0^2, V)+\frac{4\pi(N_2-P_2)}{|S|},\label{b8}
\eer
 where and in what follows we use the notation
 \ber
  f(s^1, s^2, t)\equiv\frac{\re^{s^1-s^2+t}-1}{\re^{s^1-s^2+t}+1}=\frac{\re^{s^1+t}-\re^{s^2}}{\re^{s^1+t}+\re^{s^2}},
\quad s^1,s^2,t\in\bfR. \label{b9}
 \eer
For fixed $s^1, s^2\in \mathbb{R}$, we have
 \ber
  0<\frac{\ud}{\ud t}f(s^1, s^2,  t)=\frac{2\re^{s^1+t}\re^{s^2}}{(\re^{s^1+t}+\re^{s^2})^2}\le \frac12, \quad \forall\,t\in\mathbb{R}.\label{e43}
 \eer

We now  show that the condition consisting of \eqref{C1} and \eqref{C2} is  necessary for the existence of solutions for \eqref{b7}--\eqref{b8}. In fact, integrating \eqref{b7}--\eqref{b8} over $S$,  we find
 \ber
  \int_Sf(u_0^1, u_0^2, U)\,\ud\Omega_g&=&a|S|, \label{b10}\\
  \int_Sf(v_0^1, v_0^2, V)\,\ud\Omega_g&=&b|S|, \label{b11}
 \eer
where $a, b$ are constants  defined by
 \ber
  a&\equiv&-\frac{\pi}{|S|}\left(N_1-P_1+N_2-P_2\right),  \label{b12}\\
   b&\equiv&-\frac{\pi}{|S|}\left(N_1-P_1+2(N_2-P_2)\right). \label{b13}
 \eer

 From \eqref{b10}--\eqref{b11} we see that the quantized integrals \eqref{b4a}--\eqref{b4b} hold.

On the other hand, noting
\be
-1<  f(s^1, s^2, t)< 1 \quad \text{ for any}\quad s^1, s^2, t\in \mathbb{R}, \label{b13a}
\ee
we arrive at
\ber
 |a|<1\quad \text{and}\quad |b|<1,  \label{b14}
\eer
which is equivalent to \eqref{C1} and \eqref{C2}. Thus the inequalities  \eqref{C1} and \eqref{C2} are necessary for a solution to exist.

\section{Proof of existence for the Tong--Wong system}
\setcounter{equation}{0}

In this section we establish Theorem \ref{thm2.1} or Theorem \ref{thm3.1} through a thorough study of the coupled
vortex equations (\ref{x3.2}) and (\ref{x3.3}).  To this end, we  recast the problem into a variational problem and  apply a direct minimization  approach recently developed in \cite{LiebYang}.

We have shown that the condition   \eqref{z0} is necessary to the existence of a solution to  \eqref{x3.2}--\eqref{x3.3} on $S$,
and in what follows we prove that it is also sufficient.

To formulate the problem into a variational structure, we set
\be
 f=U,\quad h=U+2V,\quad \text{or}\quad U=f,\quad  V=\frac{h-f}{2}.
\ee
The we rewrite the equations  \eqref{z3}--\eqref{z4}  equivalently as
 \ber
 \Delta f&=&4\left(\re^{u_0^1+f}-1\right)-2\left(\re^{v_0^1+\frac{h-f}{2}}-1\right)+\frac{4\pi N_1}{|S|}, \label{z5}\\
 \Delta h&=&2\left(\re^{v_0^1+\frac{h-f}{2}}-1\right)+\frac{4\pi (N_1+2N_2)}{|S|}.\label{z6}
\eer

Then we directly check  that the equations \eqref{z5}--\eqref{z6} are the Euler--Lagrange equations of the following functional
 \ber
  I(f, h)&=&\frac12\left(\|\nabla f\|_2^2+\|\nabla h\|_2^2\right)+4\int_S\left(\re^{u_0^1+f}-f+\re^{v_0^1+\frac{h-f}{2}}-\frac{h-f}{2}\right)\ud\Omega_g\nn\\
   && +\frac{4\pi N_1}{|S|}\int_S f\ud\Omega_g+\frac{4\pi (N_1+2N_2)}{|S|}\int_S h\ud\Omega_g. \label{z7}
 \eer
 Here and in what follows we use the following notation
 \ber
 \|\nabla w\|_2^2=\int_S|\nabla w|^2\ud \Omega_g\equiv  \int_Sg^{jk}\partial_jw\partial_kw\ud \Omega_g.\label{zz}
 \eer

We know that the Sobolev space $W^{1, 2}(S)$ (cf. \cite{Aubin}) can be decomposed as
$
W^{1, 2}(S)=\dot{W}^{1, 2}(S)\oplus\mathbb{R}$
where
\ber
\dot{W}^{1, 2}(S)\equiv \left\{w\in W^{1, 2}(S)\Big|\quad \int_S w\,\ud\Omega_g=0 \right\} \label{b18}
\eer
is a closed subspace of $W^{1, 2}(S)$.

To save notation, in the following of this paper we also  use $W^{1, 2}(S), \dot{W}^{1, 2}(S)$ and $L^p(S)$ to denote the spaces of  vector-valued functions.

For  $f, h\in W^{1, 2}(S)$ we decompose them as
 \be
  f=f'+\overline{f}, \, h=h'+\overline{h}, \quad f', h'\in \dot{W}^{1,2}(S), \quad \overline{f}, \overline{h}\in \mathbb{R}.\label{zz*}
 \ee

 On the subspace  $\dot{W}^{1,2}(S)$ there hold the Poincar\'{e} inequality
 \ber
   \int_S w^2\ud \Omega_g\equiv\|w\|_2^2\le C\|\nabla w\|_2^2,\quad w\in \dot{W}^{1,2}(S)\label{zp}
 \eer
 and    the  Moser--Trudinger inequality \cite{Aubin,font}
 \be
  \int_S \re^w\ud \Omega_g\le C \exp\left(\frac{1}{16\pi}\int_S|\nabla w|^2\ud x\right),\quad w\in \dot{W}^{1,2}(S), \label{z7'}
 \ee
 where $C$  is a generic positive constant.
We see from \eqref{z7'} that the functional $I$ defined by \eqref{z7} is a $C^1$-functional.

By the definition of $I$ and the  decomposition \eqref{zz*} we have
\ber
 I(f,h)&=&\frac12(\|\nabla f'\|_2^2+\|\nabla h'\|_2^2)+ 4\left(\re^{\overline{f}}\int_S\re^{u_0^1+f'}\ud\Omega_g-a_1\overline{f}\right)\nn\\
 &&+ 4\left(\re^{ \frac{\overline{h}-\overline{f}}{2}}\int_S\re^{v_0^1+\frac{h'-f'}{2}}\ud\Omega_g- a_2\frac{[\overline{h}-\overline{f}]}{2}\right),\label{z9}
\eer
where $a_1,a_2$,  defined by \eqref{zz1}--\eqref{zz2}, are positive, as ensured by \eqref{z0}.

Hence, by \eqref{z9} and the   Jensen  inequality, we obtain
\ber
 &&I(f,h)-\frac12(\|\nabla f'\|_2^2+\|\nabla h'\|_2^2)\ge4 \left(|S|\re^{\overline{f}}-a_1\overline{f}
 + |S|\re^{ \frac{\overline{h}-\overline{f}}{2}}-a_2\frac{[\overline{h}-\overline{f}]}{2}\right). \label{z10}
\eer

From \eqref{z9}--\eqref{z10} we also see that
\ber
 I(f, h)\ge 4\left(\ln\frac{|S|}{a_1}+\ln\frac{|S|}{a_2}\right),\label{z11}
\eer
which implies  the functional $I$  is bounded from below and the minimization problem
\be
 a_0\equiv\min \left\{I(f, h)\big|\, (f, h)\in W^{1, 2}(S)\right\}
\ee
is well-defined.

Let $\{(f_k, h_k)\}$ be a minimizing sequence.  Noting that  the function $m(t)=\alpha\re^t-\beta, \, \alpha,\beta>0$ satisfies $m(t)\to +\infty$ as
$t\to \pm\infty$, we see from \eqref{z10} that $\overline{f}_k,\frac{ \overline{h}_k-\overline{f}_k}{2}$ must be bounded for all $k$, which implies  $\{(\overline{f}_k, \overline{h}_k)\}$  are  bounded for all $k$.  And  \eqref{z10} also implies
$\{(\nabla f_k'\,,\nabla h_k')\}$ are bounded in $L^2(S)$ for all $k$. Then by the Poincar\'{e} inequality \eqref{zp}, we conclude that $\{(f_k', h_k')\}$ are  bounded in $\dot{W}^{1, 2}(S)$, which
with the boundedness of $\{(\overline{f}_k, \overline{h}_k)\}$  imply that $\{(f_k,  h_k)\}$ are bounded in $W^{1,2}(S)$ for all $k$.  Hence, there exits a subsequence of
$\{(f_k, h_k)\}$, still denoted by $\{(f_k, h_k)\}$, such that $\{(f_k, h_k)\}$ converges weakly to some $(f_\infty, h_\infty)\in W^{1,2}(S)$.

It is easy to see that the functional $I$ is also weakly lower semi-continuous. Then the limit $(f_\infty, h_\infty)\in W^{1,2}(S)$  is a critical point of
$I$. Of course, it gives a solution for the system \eqref{z5}--\eqref{z6}, and hence for \eqref{z3}--\eqref{z4}. So the sufficiency of \eqref{z0}   follows.

We directly see that the functional $I$ is strictly convex. Therefore the  functional  $I$  admits at most one critical point. That is to say, a solution of  \eqref{z3}--\eqref{z4} must be  unique.

Therefore we have completed the proof of Theorem \ref{thm3.1}.

\section{Proof of existence for the vortex and anti-vortex system}
\setcounter{equation}{0}

In this section, we prove that the condition comprised of \eqref{C1} and \eqref{C2} is also sufficient for the existence of a solution of
the coupled equations \eqref{b1} and \eqref{b2}.  We will extend a fixed-point theorem argument used in \cite{yang1} when treating a single equation.

To do so, it is convenient to rewrite the equations \eqref{b7} and \eqref{b8}  equivalently as
 \ber
 \Delta U&=&8\left(f(u_0^1, u_0^2, U)-a\right)-4\left(f(v_0^1, v_0^2, V)-b\right), \label{b15}\\
 \Delta V&=&-4\left(f(u_0^1, u_0^2, U)-a\right)+4\left(f(v_0^1, v_0^2, V)-b\right), \label{b16}
\eer
where $a, b$ are  defined by \eqref{b12}--\eqref{b13}.

We begin with the following lemma.
\begin{lemma}\label{lemb1}
For any $(U', V')\in \dot{W}^{1, 2}(S)$, there exists a unique pair $(c_1(U'), c_2(V'))\in \mathbb{R}^2$ such that
\ber
 \int_Sf(u_0^1, u_0^2, U'+c_1(U'))\,\ud\Omega_g&=&a|S|,\label{b19}\\
 \int_Sf(v_0^1, v_0^2, V'+c_2(V'))\,\ud\Omega_g&=&b|S|,\label{b20}
\eer
where $a, b$ are defined by \eqref{b12}--\eqref{b13}.
\end{lemma}

{\bf Proof.} Under the condition consisting of  \eqref{C1} and \eqref{C2}, we easily  see that
 \ber
  -1<a,b<1.\label{b21}
 \eer
Noting  the expression \eqref{b9},  for any $(U',V')\in \dot{W}^{1, 2}(S)$, we have
 \ber
  \int_Sf(u_0^1, u_0^2, U'+t)\ud\Omega_g,  \int_Sf(v_0^1, v_0^2, V'+t)\ud\Omega_g\to |S|\quad \text{as}\quad t\to \infty\label{b22}
 \eer
and
 \ber
  \int_Sf(u_0^1, u_0^2, U'+t)\ud\Omega_g,  \int_Sf(v_0^1, v_0^2, V'+t)\ud\Omega_g\to-|S|\quad \text{as}\quad t\to-\infty.\label{b23}
 \eer

 Then, for any $(U', V')\in \dot{W}^{1, 2}(S)$,  we conclude from \eqref{b21}, \eqref{b22} and \eqref{b23} that there exists a
point $(c_1(U'), c_2(V'))\in \mathbb{R}^2$  such that \eqref{b19} and \eqref{b20} hold.

 The uniqueness of $(c_1(U'),   c_2(V'))$ follows from the strict monotonicity of $ f(s^1, s^2,t)$ with respect to $t$ (see \eqref{e43}).

\begin{lemma}\label{lemb2}
For any $(U',  V')\in \dot{W}^{1, 2}(S)$, let  $(c_1(U'), c_2(V'))$ be defined in Lemma \ref{lemb1}.  Then, the mapping
 $(c_1(\cdot), c_2(\cdot)): \dot{W}^{1, 2}(S)\to \mathbb{R}^2$,   is continuous with respect to the weak topology of $\dot{W}^{1, 2}(S)$.
\end{lemma}

{\bf Proof. } Take a weakly convergent sequence  $\{(U_k', V_k')\}$ in $\dot{W}^{1, 2}(S)$ such that  $(U_k', V_k')\to (U_0', V_0')$ weakly in $\dot{W}^{1, 2}(S)$.
  Then we see that
   \ber
   (U_k', V_k')\to (U_0', V_0')  \quad \text{strongly in}\quad  L^p(S)  \quad \text{for any}\quad  p\ge1,\label{b26'}
  \eer
by the compact embedding $W^{1, 2}(S)\hookrightarrow L^p(S)(p\ge1)$.
We aim to prove that $(c_1(U_k'), c_2(V_k'))\to (c_1(U_0'), c_2(V_0'))$ as $k\to \infty$.

Claim: The sequence $\{(c_1(U_k'), c_2(V_k'))\}$  is bounded.

To  show this claim we first prove that  $\{(c_1(U_k'), c_2(V_k'))\}$ is bounded from above. We argue by contradiction. Without loss of generality, assume $c_1(U_k')\to \infty$ as $k\to\infty$.
Noting \eqref{b26'} and using the Egorov theorem, we see that for any $\vep>0$, there is a large constant $K_\vep>0$ and a subset $S_\vep\subset S$ such that
\ber
 |U_k'|\le K_\vep, \quad x\in S\setminus S_\vep, \quad |S_\vep|<\vep,\quad \forall\, k.\label{b23a}
\eer

Then by  \eqref{b23a} and \eqref{b13a}  we have
\ber
|a||S|&=&\left|\int_{S\setminus S_\vep}f(u_0^1, u_0^2, U_k'+c_1(U_k'))\ud\Omega_g+\int_{S_\vep}f(u_0^1, u_0^2, U_k'+c_1(U_k'))\ud\Omega_g\right|\nn\\
  &\ge&\left|\int_{S\setminus S_\vep}f(u_0^1, u_0^2, U_k'+c_1(U_k'))\ud\Omega_g\right|-\left|\int_{S_\vep}f(u_0^1, u_0^2, U_k'+c_1(U_k'))\ud\Omega_g\right|\nn\\
  &\ge&\int_{S\setminus S_\vep}f(u_0^1, u_0^2, c_1(U_k')-K_\vep)\ud\Omega_g-\vep.\label{b24}
\eer
Hence taking $k\to \infty$ in \eqref{b24} we get
\berr
 |a||S|\ge |S\setminus S_\vep|-\vep\ge |S|-2\vep.
\eerr
Noting that $\vep$ is arbitrary, we obtain
\[|a|\ge1,\]
which contradicts the condition \eqref{C1} $(|a|<1)$.  Hence  the sequence  $\{(c_1(U_k'), c_2(V_k'))\}$ is bounded from above.

Now we   show that $\{(c_1(U_k'), c_2(V_k'))\}$ is also  bounded from below.  In fact, we may suppose $c_1(U_k')\to -\infty$ as $k\to\infty$.
Using \eqref{b23a} and \eqref{b13a}, we have
\ber
 a|S|&=&\int_{S\setminus S_\vep}f(u_0^1, u_0^2, U_k'+c_1(U_k'))\ud\Omega_g+\int_{S_\vep}f(u_0^1, u_0^2, U_k'+c_1(U_k'))\ud\Omega_g\nn\\
  &\le&\int_{S\setminus S_\vep}f(u_0^1, u_0^2, c_1(U_k')+K_\vep)\ud\Omega_g+\vep.\label{b24a}
\eer
Then letting $k\to \infty$ in \eqref{b24a}, we obtain
\berr
 a|S|\le -|S\setminus S_\vep|+\vep\le -|S|+2\vep,
\eerr
which implies $a\le-1$ since $\vep>0$ is arbitrary. Hence we get a  contradiction with   the condition \eqref{C1} again. So the sequence  $\{(c_1(U_k'), c_2(V_k'))\}$ is bounded from below.
Therefore  the claim follows.

By the claim above, up to a subsequence, we may assume that
 \ber
  (c_1(U_k'), c_2(V_k'))\to (c_1', c_2') \quad \text{as}\quad k\to \infty \quad \text{for some}\quad  (c_1', c_2')\in\mathbb{R}^2. \label{b24z}
 \eer

Then, using \eqref{e43}, the Schwartz inequality,  \eqref{b26'} and \eqref{b24z} we have
\ber
&&\left|\int_Sf(u_0^1, u_0^2, U_k'+c_1(U_k'))\,\ud\Omega_g-\int_Sf(u_0^1, u_0^2, U_0'+c_1')\,\ud\Omega_g\right|\nn\\
&&=\left|\int_Sf_t(u_0^1, u_0^2, \theta[U_k'+c_1(U_k')]+[1-\theta][U_0'+c_1'])(U_k'-U_0'+c_1(U_k')-c_1')\,\ud\Omega_g\right|
\nn\\
&&\le\frac12\int_S\left|U_k'-U_0'+c_1(U_k')-c_1'\right|\,\ud\Omega_g\nn\\
&&\le \frac12\left(|S|^{\frac12}\|U_k'-U_0'\|_2+|S||c_1(U_k')-c_1'|\right)\to 0 \quad \text{as}\quad k\to \infty, \label{nn1}
\eer
where $\theta\in(0,1)$.  Noting  \eqref{nn1} and
\be
\int_Sf(u_0^1, u_0^2, U_k'+c_1(U_k'))\,\ud\Omega_g=a|S|,
\ee
we have
 \ber
 \int_Sf(u_0^1, u_0^2, U_0'+c_1')\,\ud\Omega_g=a|S|.
 \eer

 Similarly, we get\be
\int_Sf(v_0^1, v_0^2, V_0'+c_2')\,\ud\Omega_g=b|S|.
\ee

Hence  from Lemma \ref{lemb1}  we see that $(c_1', c_2')=(c_1(U_0'), c_2(V_0'))$. Then Lemma \ref{lemb2} follows.

At this point we can define an operator
\berr
 T:\dot{W}^{1, 2}(S)\to  \dot{W}^{1, 2}(S)
\eerr
as follows. For $(U', V')\in  \dot{W}^{1, 2}(S)$, let $(c_1(U'), c_2(V'))$ be defined by Lemma \ref{lemb1}.
Define $(\tilde{U}',\tilde{V}')=T(U',V')$ where $\tilde{U}'$ and $\tilde{V}'$ are the unique solutions of
\ber
 \Delta \tilde{U}'&=& 8\left(f(u_0^1, u_0^2, U'+c_1(U'))-a\right)-4\left(f(v_0^1, v_0^2, V'+c_2(V'))-b\right), \label{b25}\\
 \Delta \tilde{V}'&=&-4\left(f(u_0^1, u_0^2, U'+c_1(U'))-a\right)+4\left(f(v_0^1, v_0^2, V'+c_2(V'))-b\right), \label{b26}
\eer
respectively.
In fact,    for any $(U', V')\in \dot{W}^{1, 2}(S)$, since  the right-hand sides of \eqref{b25} and \eqref{b26} have zero averages,  the
solutions $\tilde{U}'$ and $\tilde{V}'$ of   \eqref{b25} and \eqref{b26}, respectively, are unique (cf. \cite{Aubin}).

Next we show that the operator $T$ admits a fixed point in $\dot{W}^{1, 2}(S)$.  To this end, we first  establish the following lemma.
\begin{lemma}\label{lemb3}
The above operator $T:\dot{W}^{1, 2}(S)\to  \dot{W}^{1, 2}(S)$ is  completely continuous.
\end{lemma}

{\bf Proof.} Assume $(U_k', V_k')\to (U_0', V_0')$ weakly in $\dot{W}^{1, 2}(S)$.
 Hence by the compact embedding theorem we see that \eqref{b26'} holds.

 Denote
   \ber
     (\tilde{U}'_k,\tilde{V}'_k)= T(U'_k, V'_k) \quad \text{and}\quad  (\tilde{U}'_0,\tilde{V}'_0)= T(U'_0, V'_0).\label{b27}
   \eer
Therefore we have
 \ber
 \Delta (\tilde{U}'_k-\tilde{U}'_0)&=&8\big(f(u_0^1, u_0^2, U'_k+c_1(U'_k))-f(u_0^1, u_0^2, U'_0+c_1(U'_0))\big)\nn\\
 &&-4\big(f(v_0^1, v_0^2, V'_k+c_2(V'_k))-f(v_0^1, v_0^2, V'_0+c_2(V'_0))\big)\nn\\
 &=&8f_t(u_0^1, u_0^2, \hat{U}'+\hat{c}_1)\big(U_k'-U_0'+c_1(U_k')-c_1(U_0')\big)\nn\\
 &&-4f_t(v_0^1, v_0^2, \hat{V}'+\hat{c}_2)\big(V_k'-V_0'+c_2(V_k')-c_2(V_0')\big), \label{b28}\\[2mm]
 \Delta (\tilde{V}'_k-\tilde{V}'_0)&=&-4\big(f(u_0^1, u_0^2, U'_k+c_1(U'_k))-f(u_0^1, u_0^2, U'_0+c_1(U'_0))\big)\nn\\
 &&+4\big(f(v_0^1, v_0^2, V'_k+c_2(V'_k))-f(v_0^1, v_0^2, V'_0+c_2(V'_0))\big)\nn\\
 &=&-4f_t(u_0^1, u_0^2, \hat{U}'+\hat{c}_1)\big(U_k'-U_0'+c_1(U_k')-c_1(U_0')\big)\nn\\
 &&+4f_t(v_0^1, v_0^2, \hat{V}'+\hat{c}_2)\big(V_k'-V_0'+c_2(V_k')-c_2(V_0')\big),\label{b29}
 \eer
 where $\hat{U}'_k$ lies between $U'_k$ and $U'_0$, $\hat{V}'_k$ between $V'_k$ and $V'_0$, $\hat{c}_1$ between $c_1(U_k')$ and $c_1(U_0')$,
and $\hat{c}_2$ between $c_2(V_k')$ and $c_2(V_0')$.

 Multiplying both sides of \eqref{b28} and \eqref{b29} by $\tilde{U}'_k-\tilde{U}'_0$ and $\tilde{V}'_k-\tilde{V}'_0$, respectively,   and integrating by parts, we obtain
 \ber
  \|\nabla (\tilde{U}'_k-\tilde{U}'_0)\|_2^2&\le&\int_S \Big\{4\big(|U_k'-U_0'|+|c_1(U_k')-c_1(U_0')|\big) \nn\\
  &&+2\big(|V_k'-V_0'|+|c_2(V_k')-c_2(V_0')|\big)\Big\}|\tilde{U}'_k-\tilde{U}'_0|\,\ud\Omega_g,\label{b30}\\
    \|\nabla (\tilde{V}'_k-\tilde{V}'_0)\|_2^2&\le&2\int_S \Big(|U_k'-U_0'|+|c_1(U_k')-c_1(U_0')| \nn\\
  &&+|V_k'-V_0'|+|c_2(V_k')-c_2(V_0')|\Big)|\tilde{V}'_k-\tilde{V}'_0|\,\ud\Omega_g,\label{b31}
 \eer
 where  the property \eqref{e43} is used.

Combining \eqref{b30} with \eqref{b31}, and using the Poincar\'{e} inequality, we arrive at
 \ber
  \|\nabla (\tilde{U}'_k-\tilde{U}'_0)\|_2^2+\|\nabla (\tilde{V}'_k-\tilde{V}'_0)\|_2^2&\le& C\Big(\|U_k'-U_0'\|_2^2+\|V_k'-V_0'\|_2^2\nn\\
  &&+|c_1(U_k')-c_1(U_0')|^2+|c_2(V_k')-c_2(V_0')|^2\Big)\label{b32}
 \eer
 for some $C>0$.
Then, from  \eqref{b26'}, Lemma \ref{lemb2}, and \eqref{b32}, we see that
\[(\nabla \tilde{U}'_k,\nabla \tilde{V}'_k) \to (\nabla\tilde{U}'_0,\nabla\tilde{V}'_0) \quad \text{strongly in}\quad  L^2(S)  \quad \text{as}\quad  k\to\infty,\]
which, with \eqref{b26'}, yields
\[(\tilde{U}'_k,\tilde{V}'_k) \to (\tilde{U}'_0,\tilde{V}'_0) \quad \text{strongly in}\quad  \dot{W}^{1,2}(S)  \quad \text{as}\quad  k\to\infty.\]
 Then the proof of Lemma \ref{lemb3} is complete.

Before applying the Leray--Schauder fixed-point theory, we need to estimate the solution of the fixed-point equation,
 \ber
  (U'_t, V'_t)&=&tT(U'_t, V'_t), \quad 0\le t\le 1.\label{b33}
 \eer

\begin{lemma}\label{lemb4}
For any $(U'_t, V'_t)$ satisfying \eqref{b33}, there exists a constant $C>0$ independent of $t\in[0, 1]$ such that
 \ber
  \|U'_t\|_{\dot{W}^{1, 2}(S)}+ \|V'_t\|_{\dot{W}^{1, 2}(S)}\le C.\label{b34}
 \eer
\end{lemma}

{\bf Proof.} From \eqref{b33} we  have
 \ber
 \Delta U'_t&=&8t\left(f(u_0^1, u_0^2, U'_t+c_1(U'_t))-a\right)-4t\left(f(v_0^1, v_0^2, V'_t+c_2(V'_t))-b\right), \label{b35}\\
 \Delta V'_t&=&-4t\left(f(u_0^1, u_0^2, U'_t+c_1(U'_t))-a\right)+4t\left(f(v_0^1, v_0^2, V'_t+c_2(V'_t))-b\right). \label{b36}
\eer

Multiplying both sides of \eqref{b35} and \eqref{b36} by $U_t'$ and $V_t'$, respectively, and integrating by parts, we see that
\berr
\|\nabla U_t'\|_2^2&\le&\int_S \Big(8|f(u_0^1, u_0^2, U'_t+c_1(U'_t))|+4|f(v_0^1, v_0^2, V'_t+c_2(V'_t))|\Big)|U'_t|\ud\Omega_g\nn\\
 &\le& 12\int_S|U'_t|\ud\Omega_g,\\
\|\nabla V_t'\|_2^2&\le&\int_S \Big(4|f(u_0^1, u_0^2, U'_t+c_1(U'_t))|+4|f(v_0^1, v_0^2, V'_t+c_2(V'_t))|\Big)|V'_t|\ud\Omega_g\nn\\
&\le&  8\int_S|V'_t|\ud\Omega_g,
\eerr
where we have used \eqref{b13a}.
Then  by the  Poincar\'{e} inequality, we get the desired estimate \eqref{b34}.

Now using Lemmas \ref{lemb3}, \ref{lemb4}, and the Leray--Schauder fixed-point theorem (cf. \cite{GL}), we see that the operator $T$ admits a fixed point, say $(U', V')$, in  $\dot{W}^{1, 2}(S)$.
Thus  $(U'+c_1(U'), V'+c_2(V'))$ is a solution of  \eqref{b15} and \eqref{b16},  i.e. a solution of \eqref{b7} and \eqref{b8}.

Hence  we have completed the proof  of Theorem \ref{thb1}.

\section{Explicit calculation of minimum energy}
\setcounter{equation}{0}

In this section we establish the minimum energy formula (\ref{xE}) and show how it is stratified topologically.

By the equations  \eqref{x1.17}--\eqref{x1.20}, the fact $*1=\dd\Omega_g$, and \eqref{b4a}--\eqref{b4b}, we see that
 \ber
  \int_S(\hat{F}-\tilde{F})&=&4\int_S*\frac{1-\re^u}{\re^u+1}-2\int_S*\frac{1-\re^v}{\re^v+1}=2\pi(N_1-P_1), \label{c1}\\
  \int_S\tilde{F}&=&-2\int_S*\frac{1-\re^u}{\re^u+1}+2\int_S*\frac{1-\re^v}{\re^v+1}=2\pi(N_2-P_2), \label{c2}
 \eer
are valid, which give us
\ber
\int_S\hat{F}=2\pi(N_1-P_1+N_2-P_2).  \label{c3}
\eer

To calculate the lower bound of the energy, we need to compute the fluxes contributed by the current densities
$K(q)$ and $K(p)$.

Take a coordinate chart $\{\mathcal{U}_j\}$ of $S$. Assume $z_{1,j}''\in \mathcal{U}_j$, $j=1,\dots, P_1$. In local
coordinates, we have
 $D_iq=\partial_iq-\ri(\hat{A}_i-\tilde{A}_i)q$, $i=1, 2$ and the density $K(q)$ in $\mathcal{U}_j$ can be written as
 \ber
  K(q)=-\frac{2|q|^2}{1+|q|^2}(\hat{F}-\tilde{F})+\ri\frac{D_iq\overline{D_jq}-\overline{D_i}qD_jq}{(1+|q|^2)^2}\ud x^i\wedge\ud x^j. \label{c4}
 \eer
Besides, in
$
K(q)=\ud J(q)$, we have
\ber
 \quad J(q)=\frac{\ri}{1+|q|^2}(q\overline{D_iq}-\overline{q}D_iq)\ud x^i.\label{c5}
\eer

Then  it follows from the Stokes  formula that
\ber
 \int_SK(q)=\int_S\ud J(q)=\sum\limits_{j=1}^{P_1}\lim\limits_{r\to0}\oint_{\partial B(z_{1,j}'', r)} J(q),\label{c6}
\eer
where $B(z, r)$  denotes  a disc  centered at $z$  with radius $r>0$ and all the line integrals are   taken
 counterclockwise.

 Note that near $z_{1,j}''\in \mathcal{P}(q)$,  the section $q$ has the representation
  \ber
   q(z)=z^{-1}h_j(z,\overline{z}), \quad z=x^1+\ri x^2, \quad x^1(z_{1,j}'')=x^2(z_{1,j}'')=0,\label{c7}
   \eer
 where $h_j$ is a non-vanishing function defined near $z_{1,j}''$.

 From the equation  \eqref{x1.17} we  see that
  \ber
  \hat{A}_1-\tilde{A}_1=-2{\rm Re}(\ri\overline{\partial}\ln u), \quad \hat{A}_2-\tilde{A}_2=-2{\rm Im}(\ri\overline{\partial}\ln u),\label{c8}
  \eer
which, with $u=\ln|q|^2$, implies
\ber
 D_1q=(\partial+\overline{\partial})q+\left(\frac{\partial\overline{q}}{\overline{q}}-\frac{\overline{\partial}q}{q}\right)q=q\partial u, \label{c9}\\
 D_2q=\ri(\partial-\overline{\partial})q+\ri\left(\frac{\overline{\partial}q}{q}+\frac{\partial\overline{q}}{\overline{q}}\right)q=\ri q\partial u.\label{c10}
\eer

Then,  by   \eqref{c6},  \eqref{c9}, and \eqref{c10},  we have
\ber
 \oint_{\partial B(z_{1,j}'', r)} J(q)&=&\ri\oint_{\partial B(z_{1,j}'', r)} \frac{|q|^2}{1+|q|^2}(\overline{[\partial}-\partial]u\ud x^1-\ri[\overline{\partial}+\partial]u\ud x^2)\nn\\
 &=&\oint_{\partial B(z_{1,j}'', r)} \frac{\re^u}{1+\re^u}(\partial_2u\ud x^1-\partial_1u \ud x^2).\label{c11}
\eer

Noting \eqref{c7},  near $z_{1,j}''\in \mathcal{P}(q)$,  we see that
 \ber
 u=-2\ln|z|+w_j, \label{c12}
 \eer
where $w_j$ is a smooth function.  Thus  we obtain
\ber
 \lim\limits_{r\to0}\oint_{\partial B(z_{1,j}'', r)} J(q)=4\pi,\label{c13}
\eer
which, with \eqref{c6},  gives
\ber
 \int_SK(q)=4\pi P_1.\label{c14}
\eer

 Following a similar procedure, we have
 \ber
 \int_SK(p)=4\pi P_2.\label{c15}
\eer

As described in \cite{SSY}, the normalized integrals $\frac1{4\pi}\int  K(q)$ and $\frac1{4\pi}\int K(p)$, counting the numbers $P_1,P_2$ of
anti-vortices of the two species, are the Thom classes of the dual bundle $L^*\to S$, of two respective classification (Chern) classes,
$\frac1{2\pi}\int(\hat{F}-\tilde{F})$ and $\frac1{2\pi}\int \tilde{F}$.

Hence, by \eqref{x1.13}--\eqref{x1.15}, \eqref{c3},  \eqref{c14}, and \eqref{c15}, we obtain the following topologically
stratified minimum energy
\ber
E=\int_S2([\hat{F}-\tilde{F}]+\tilde{F}+K(q)+K(p))=4\pi(N_1+P_1+N_2+P_2),
\eer
as stated in Theorem \ref{mainthm}.

\section{Conclusions and remarks}
\setcounter{equation}{0}

In this work we have extended the formalism of Tong and Wong \cite{TW} of a product Abelian Higgs theory describing a coupled vortex system with magnetic impurities to accommodate
coexisting vortices and anti-vortices of two species realized as topological solitons governed by a BPS system of equations. In additional to the usual first Chern classes suited over a
complex Hermitian line bundle, the presence of  anti-vortices switches on the Thom classes over the dual bundle, as in \cite{SSY}. When the underlying Riemann surface $S$ where vortices
and anti-vortices reside is compact, we have established a theorem which spells out a necessary and sufficient condition,
consisting of two inequalities, (\ref{C1}) and (\ref{C2}), for prescribed  $N_1, N_2$ vortices and $P_1,P_2$ anti-vortices,
of two respective species, to exist.

This necessary and sufficient condition contains a few special situations worthy of mentioning.

\begin{enumerate}
\item[(i)]
When $N_2=P_2=0$ (only vortices and anti-vortices of the first species are present), the condition becomes
\be
\left|N_1-P_1\right|<\frac{|S|}\pi.
\ee

\item[(ii)] When $N_1=P_1=0$ (only vortices and anti-vortices of the second species are present), the condition reads
\be
\left|N_2-P_2\right|<\frac{|S|}{2\pi}.
\ee

\item[(iii)] When $N_1=N_2=N$ and $P_1=P_2=P$ (there are equal numbers of vortices and anti-vortices, respectively, of two species), the condition is
\be
\left|N-P\right|<\frac{|S|}{3\pi}.
\ee
\end{enumerate}

In all these situations, the numbers of vortices and anti-vortices may be arbitrarily large, provided that the differences of these numbers are kept in suitable ranges as given.

Although the vortices and anti-vortices of the two species do not appear in the model in a symmetric manner as seen in the field-theoretical Lagrangian density  and the governing
equations, they make equal contributions to the total  topologically stratified minimum energy as stated in (\ref{xE}) of an elegant form.

Let ${\cal M}(N_1,P_1,N_2,P_2)$ denote the moduli space of solutions of the BPS equations (\ref{x1.17})--(\ref{x1.20}) with $N_1+N_2$ and
$P_1+P_2$ prescribed vortices and anti-vortices, of two respective species. Since these solutions depend on at least $2(N_1+N_2+P_1+P_2)$
continuous parameters which specify the locations of zeros and poles of the two sections $q,p$, respectively, we obtain the following
lower bound for the dimensionality of ${\cal M}(N_1,P_1,N_2,P_2)$:
\be \label{6.5}
\dim({\cal M}(N_1,P_1,N_2,P_2))\geq 2(N_1+N_2+P_1+P_2).
\ee
Since we have not established the uniqueness of a solution with $N_1+N_2$ and $P_1+P_2$ prescribed vortices and anti-vortices of the two
species yet, we do not know whether the inequality (\ref{6.5}) is actually an equality. In this regard, it will be interesting to carry out
an investigation along the (well-known classical) index theory work of Atiyah, Hitchin, and Singer \cite{AHS1,AHS2}
on the Yang--Mills instantons, of Weinberg \cite{Wein}
on the BPS system of the Abelian Higgs model,  and of Lee \cite{Lee}
on supersymmetric domain walls, for our new system of equations,  (\ref{x1.17})--(\ref{x1.20}).

In a sharp contrast, if we use ${\cal M}(N_1,N_2)$ to denote the moduli space of the solutions of the Tong--Wong equations
(\ref{x2.11})--(\ref{x2.14}) with $N_1$ and $N_2$ prescribed vortices, of two respective species, the established uniqueness
of the solutions indicates the result
\be
\dim({\cal M}(N_1,N_2))= 2(N_1+N_2).
\ee
See \cite{Moore} for some recent related work.

\subsection*{Acknowledgements}
  Han was partially  supported by National Natural Science Foundation of China under Grant 11201118 and the Key Foundation for Henan colleges under Grant 15A110013. Both  authors were partially  supported by National Natural Science Foundation of China under Grants 11471100 and 11471099.

\small{

}

\end{document}